# Large out-of-plane spin-orbit torque in topological Weyl semimetal candidate TaIrTe$_4$

Lakhan Bainsla[1,2,*], Bing Zhao[1], Anamul Md. Hoque[1], Lars Sjöström[1], Nilamani Behera[3], Mahmoud Abdel-Hafiez[4], Johan Åkerman[3,5,6], Saroj P. Dash[1,7,*]

[1]*Department of Microtechnology and Nanoscience, Chalmers University of Technology, SE-41296, Göteborg, Sweden*
[2]*Department of Physics, Indian Institute of Technology Ropar, Rupnagar 140001, Punjab, India*
[3]*Department of Physics, University of Gothenburg, Göteborg, SE-41296, Göteborg, Sweden*
[4]*Department of Physics and Astronomy, Uppsala University, Box 516, SE-751 20 Uppsala, Sweden*
[5]*Center for Science and Innovation in Spintronics, Tohoku University, 2-1-1 Katahira, Aoba-ku, Sendai 980-8577 Japan*
[6]*Research Institute of Electrical Communication, Tohoku University, 2-1-1 Katahira, Aoba-ku, Sendai 980-8577 Japan*
[7]*Graphene Center, Chalmers University of Technology, SE-41296, Göteborg, Sweden*

**Abstract**

Topological quantum materials, with novel spin textures and broken crystal symmetries are suitable candidates for spintronic memory technologies. Their unique electronic properties, such as protected surface states and exotic quasiparticles, can provide an out-of-plane spin polarized current needed for external field free magnetization switching of magnets with perpendicular magnetic anisotropy. Conventional spin-orbit torque materials, such as heavy metals and topological insulators, provide only an in-plane spin polarized current, and recently explored materials with lower crystal symmetries provide very low out-of-plane spin polarized current components, which is not suitable for energy-efficient spin-orbit torque (SOT) applications. Here, we demonstrate a large out-of-plane damping-like SOT at room temperature using a topological Weyl semimetal candidate TaIrTe$_4$ with a lower crystal symmetry. We performed spin-orbit torque ferromagnetic resonance (STFMR) experiments in a TaIrTe$_4$/Ni$_{80}$Fe$_{20}$ heterostructure and observed a large out-of-plane damping-like SOT efficiency. The out-of-plane spin Hall conductivity is estimated to be $4.05\pm0.23\times10^4$ ($\hbar$/2e) ($\Omega$m)$^{-1}$, which is an order of magnitude higher than the reported values in other materials. These findings of high spin Hall conductivity and large out-of-plane SOT efficiency are suitable for the development of energy efficient and external field-free spintronic devices.

*Corresponding authors: Lakhan Bainsla, Saroj P. Dash (saroj.dash@chalmers.se)

**Introduction**

Spin-orbit torque (SOT), utilizing the charge-to-spin conversion in a high spin-orbit coupling material (SOM) to create magnetization dynamics in an adjacent ferromagnet (FM), is expected to provide a breakthrough for next-generation memory and logic technologies.[1,2] SOT-based memory devices have the potential to challenge the devices based on spin-transfer-torque, but the use of conventional SOMs leads to moderate efficiency. Furthermore, as the component of torque lies in-plane in conventional SOMs, they are only suitable for deterministic switching of in-plane magnets. [1,2] However, for thermally stable high-density memory technologies, industry requires magnets with perpendicular magnetic anisotropy (PMA), where additional measures, such as magnetic field assistance, will be required for deterministic switching. The first non-trivial requirement for a practical SOT memory technology is therefore the field-free deterministic SOT switching of FMs with PMA.[2]

To achieve this, SOMs with lower crystal symmetry are needed to generate out-of-plane damping-like torque components suitable to switch PMA ferromagnets. In contrast to conventional SOT, with a torque vector perpendicular to the plane of the electron's motion and the electric field, unconventional SOT produces a torque vector that is tilted. Recent experiments demonstrated that van der Waals SOMs with reduced crystal symmetry, such as $WTe_2$ in heterostructure with FMs, allow the generation of a non-trivial current-induced spin polarization with out-of-plane SOT symmetries.[3–8] More recently, field-free SOT switching has been reported using $WTe_2$/$Fe_3GeTe_2$ heterostructures induced by out-of-plane SOT from $WTe_2$.[3,5,9,10] However, to date, the out-of-plane SOT strengths are an order of magnitude smaller than the conventional in-plane SOT component, making it challenging to realize energy-efficient SOT devices.[3–7,9–16] Therefore, it is crucial to discover new materials that exhibit large out-of-plane spin polarization and SOT components.

The topological Weyl semimetal candidate $TaIrTe_4$ has gained significant attention as it shows the presence of bulk Weyl nodes and Fermi-arc surface states, which are unique band crossings in momentum space and novel spin textures that can give a variety of unusual electronic and charge-to-spin conversion properties.[17–21] The combination of novel topological spin textures and lower crystal symmetry hence makes $TaIrTe_4$ a promising candidate for energy-efficient SOT devices. Here, using spin-orbit torque ferromagnetic resonance (STFMR) measurements[22,23] in $TaIrTe_4$/$Ni_{80}Fe_{20}$ heterostructure, and the detailed *dc* bias and angle dependence, we demonstrate a large out-of-plane SOT, a large out-of-plane damping-like torque component, and a large spin Hall conductivity, all at room temperature.

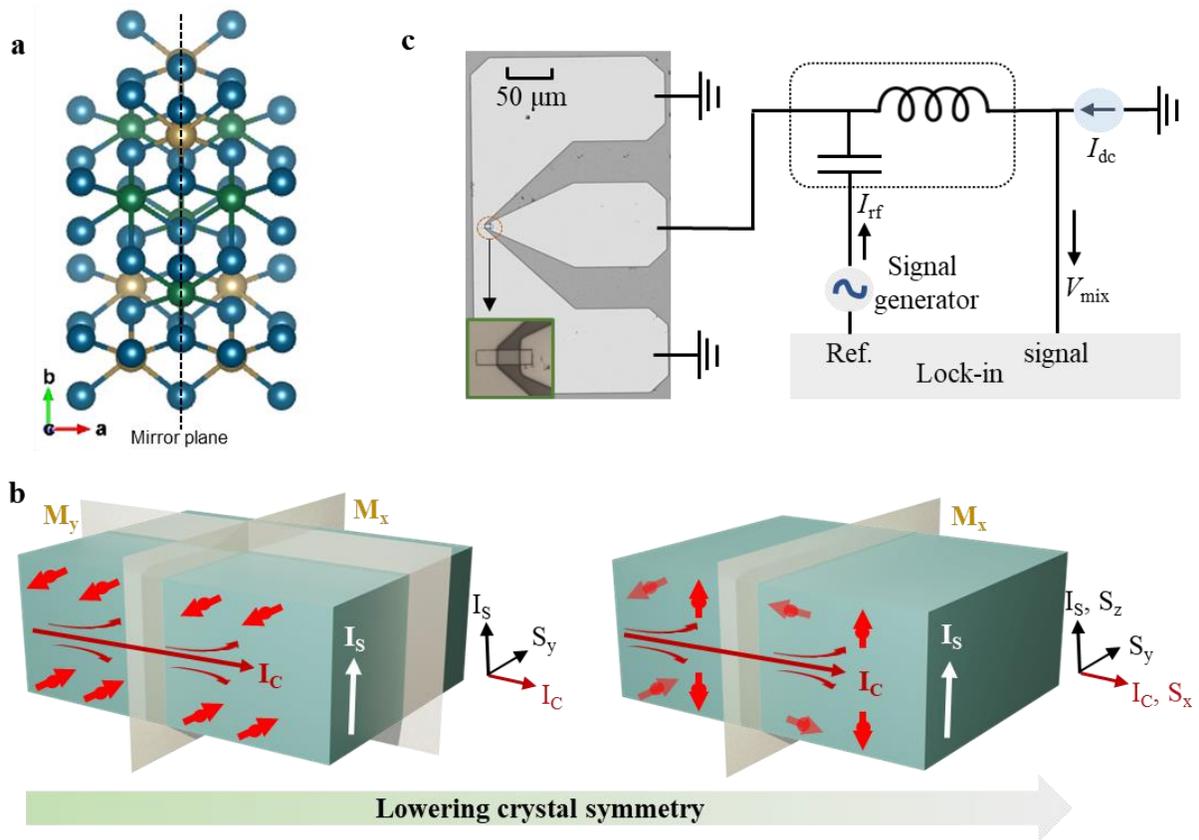

**Figure 1.** Unconventional spin-orbit torque in TaIrTe$_4$ and measurement scheme. **a,** Crystal structure of $T_d$-TaIrTe$_4$ showing lower crystal symmetry with mirror plane along a-axis only. **b,** Conventional and unconventional SOT mechanisms in higher and lower crystalline symmetry materials. In high symmetry materials, $I_C$, $I_S$ and $S$ ($S_y$) follow the conventional orthogonality relation (1$^{st}$ diagram), while one can generate unconventional out-of-plane spin currents in lower symmetry materials (2$^{nd}$ diagram; shown by removing $M_y$ mirror plane). **c,** Schematic of the spin-orbit torque ferromagnetic resonance (STFMR) measurement set-up with TaIrTe$_4$/Ni$_{80}$Fe$_{20}$ heterostructure device geometry. Scale bar is 50 μm for device image, inset shows the magnified view of the microbar device.

**Results and discussion**

TaIrTe$_4$ is a promising topological Weyl semimetal candidate due to its large spin-orbit coupling and broken crystal symmetry, exhibiting unique chiral spin textures of electronic bands for bulk Weyl nodes and Fermi-arc surface states.[24] TaIrTe$_4$ hosts only four type-II Weyl nodes providing the simplest model system with broken inversion symmetry. Additionally, the lower crystal symmetry of $T_d$-TaIrTe$_4$ structure with space group $Pmn2_1$ (Fig. **1a**), can provide unconventional charge-spin conversion due to the presence of topologically non-trivial electronic states. These properties can result in the emergence of current-induced spin polarization that does not strictly follow the orthogonal relation between the charge current ($I_C$), spin current ($I_S$), and spin polarization (S) orientation as shown in Fig. **1b**. The generation of an out-of-plane spin polarization can induce an out-of-plane SOT on the adjacent ferromagnet. In contrast to conventional in-plane SOT, which applies torque parallel to the device plane, the out-of-plane SOT can more efficiently manipulate magnetic moments with perpendicular components for high-density integration.

In this study, we focus on the unconventional charge-to-spin conversion in TaIrTe$_4$, particularly, the generation of out-of-plane SOT components. STFMR measurements are performed at room temperature to investigate the SOT in TaIrTe$_4$/Ni$_{80}$Fe$_{20}$ bilayers, Ni$_{80}$Fe$_{20}$ is also known as permalloy (Py). The schematic of the STFMR measurement setup is shown in Fig. **1c**. The STFMR microbars were fabricated using electron-beam lithography and argon ion milling. The devices were fabricated along the longer axis of the TaIrTe$_4$ flakes which is a-axis in this class of materials.[7,17,21]

In STFMR measurements, an in-plane radio frequency current $I_{rf}$ is applied along the a-axis of TaIrTe$_4$ while an in-plane magnetic field is applied at an angle $\phi$ with respect to the $I_{rf}$, as shown in Fig. **2a**. $I_{rf}$ in TaIrTe$_4$ generates a spin current in the z-direction, which is injected into the adjacent Py layer and excites the Py into a processional motion. Thanks to its anisotropic magnetoresistance (AMR), the resistance of Py oscillates with the same frequency as that of $I_{rf}$, and produces a *dc* mixing voltage V$_{mix}$, which is then measured using a lock-in amplifier. AMR is measured for a TaIrTe$_4$(133 nm)/Py(6 nm) device and a value of 0.11% is obtained as shown in Fig. **2b** and in Supplementary Fig. 2 for other devices. Figure **2c** shows the representative STFMR signals V$_{mix}$ for the TaIrTe$_4$(133 nm)/Py(6 nm) device at room temperature. The obtained V$_{mix}$ signal is then fitted using the equation,[22,25]

$$V_{mix} = SF_S(H_a) + AF_A(H_a) \qquad (1)$$

where, $F_S(H_a) = \frac{\Delta H^2}{[\Delta H^2 + (H_a - H_R)^2]}$ and $F_A(H_a) = F_S(H_a)\left[\frac{(H_a - H_R)}{\Delta H}\right]$ are symmetric and antisymmetric Lorentzian functions, respectively. *S* and *A* are the amplitudes of the symmetric $F_S$ and the antisymmetric $F_A$ signals and are proportional to the current-induced in-plane torque ($\tau_\parallel$) and out-of-plane torque ($\tau_\perp$), respectively. Here, $H_a$, $\Delta H$, and $H_R$ refer to the applied external magnetic field, the ferromagnetic resonance linewidth, and the ferromagnetic resonance field, respectively. $H_R$ and $\Delta H$ are extracted and the effective magnetization of the Py layer, $\mu_0 M_{eff.}$, is determined by fitting $f$ vs. $H_R$ to the Kittel equation, $f = \left(\frac{\gamma}{2\pi}\right)\mu_0\sqrt{(H_R - H_k)(H_R - H_k + M_{eff})}$. The effective Gilbert damping constant $\alpha$ is obtained by a linear fit of $\Delta H$ vs. $f$ using $\Delta H = \Delta H_0 + \frac{(2\pi\alpha f)}{\gamma}$.[26] The values for $\mu_0 M_{eff}$ and $\alpha$ for the TaIrTe$_4$ (90 nm)/Py (6 nm) device are given in Fig. **3a** and **3b**, respectively. The $\mu_0 M_{eff}$ and $\alpha$ values for Py, using different thicknesses of TaIrTe$_4$, are given in the Supplementary Fig. 3. The obtained values of the $\mu_0 M_{eff.}$ and $\alpha$ are comparable to literature values for Py~5 nm films.[27,28]

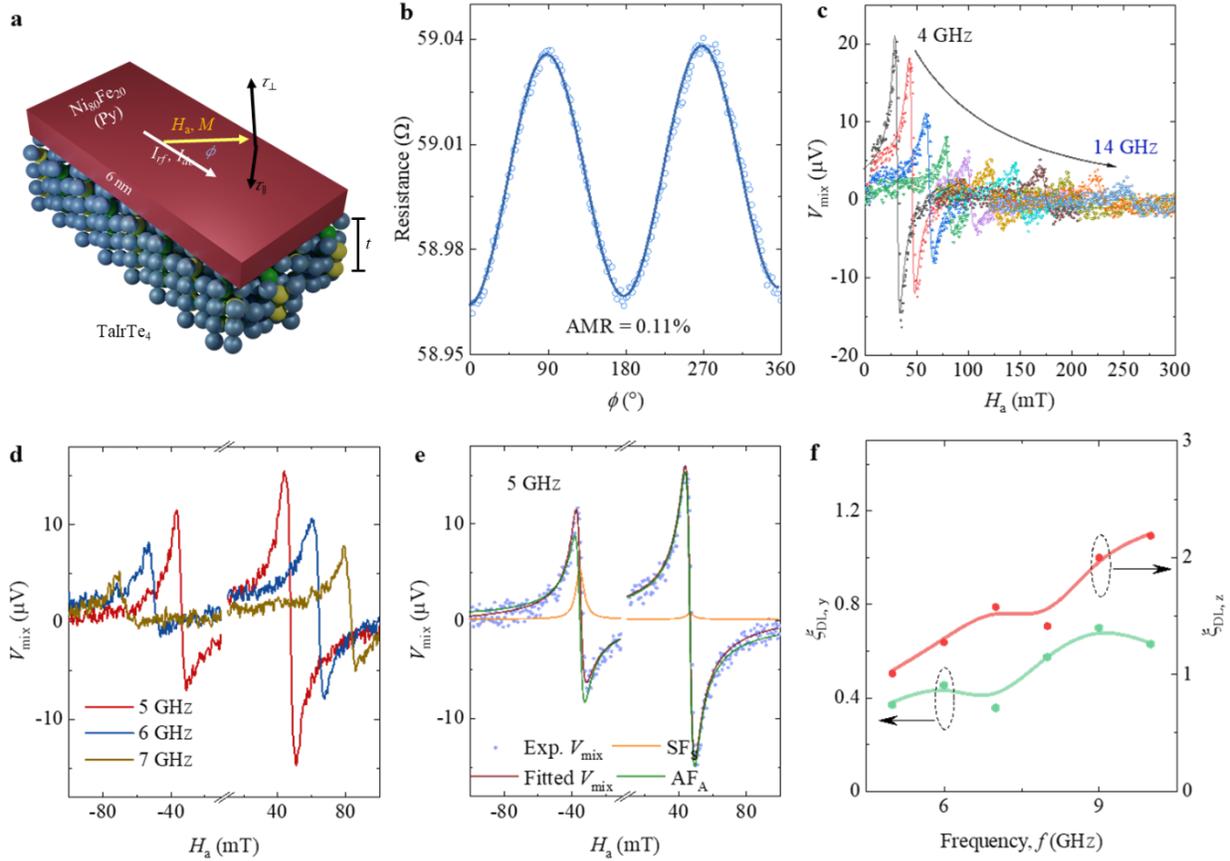

**Figure 2.** Unconventional charge-spin conversion in TaIrTe$_4$. **a,** Schematic of the TaIrTe$_4$/Py heterostructure with SOT components. **b,** Anisotropic magnetoresistance curve with an applied magnetic field of 100 mT and *dc* current of 0.6 mA. **c,** Frequency dependent STFMR spectra with in-plane magnetic field angle $\phi = 40°$ for a frequency range of 4-14 GHz. **d,** STFMR spectra with positive ($\phi$ = 40°) and negative ($\phi$ = 220°) applied magnetic field in the frequency range of 5-7 GHz. **e,** The experimental STFMR curves (Exp. $V_{mix}$), fitted curves using equation (1) (fitted $V_{mix}$), symmetric (SF$_S$) and antisymmetric (AF$_A$) contributions in the $V_{mix}$ (fitted) at 5 GHz. In **c**, and **e**, solid symbols represent the experimental STFMR data and solid lines are fit to the obtained data using Eq. (1). **f,** The evaluated y and z-polarized effective charge-to-spin conversion efficiency using lineshape analysis at different frequencies. Here, solid symbols are the estimated efficiency values and solid lines are guides to the eye. Measurements are performed in the TaIrTe$_4$(133 nm)/Py(6 nm) device.

The strengths of the current-induced torques for different $\phi$ values are related to the symmetries of the device. For example, in conventional Pt/Py bilayers, the two-fold rotational symmetry requires that the SOT changes sign when the magnetization is rotated 180° in-plane, which results in the sign reversal of $V_{mix}$ while retaining the same amplitude.[3] STFMR measurements with positive ($\phi$ = 40°) and negative ($\phi$ = 220°) applied field were performed on the TaIrTe$_4$/Py devices and a clear change in both amplitude and shape are obtained as shown in Fig. **2d** for the TaIrTe$_4$(133 nm)/Py(6 nm) device, in the frequency range of 5-7 GHz. The $V_{mix}$ signal is then fitted with Eq. (1) and the symmetric and antisymmetric components are obtained which show a large change in amplitude in Fig. **2e** for the TaIrTe$_4$(133 nm)/Py(6 nm) device at 5 GHz. It directly indicates that the SOT is affected by the reduced symmetry of the TaIrTe$_4$ layer and in particular the amplitude difference of the antisymmetric part indicates the clear presence of out-of-plane torque $\tau_\perp$.

The charge-to-spin conversion efficiency for in-plane ($\sigma_y$) and out-of-plane spin ($\sigma_z$) is evaluated using the symmetric and antisymmetric amplitudes obtained with both positive and negative applied magnetic field $H_a$ at a fixed value of $\phi$ (Supplementary Note 1).[5,22,29] The evaluated in-plane damping-like torque ($\xi_{DL,y}$) and out-of-plane damping-like torque ($\xi_{DL,z}$) efficiencies for the TaIrTe$_4$(133 nm)/Py(6 nm) device are plotted in **Fig. 2f**, which show enhancement in SOT efficiency as the $I_{rf}$ frequency is increased. Such a frequency dependent SOT efficiency has previously been observed in the WTe$_2$/NiFe system.[5] $\xi_{DL,z}$ varies from 1 to 2.19, while $\xi_{DL,y}$ varies from 0.36 to 0.63 for a frequency range of 5 to 10 GHz. However, as SOT efficiency evaluation using lineshape analysis (described in Supplementary Note 1) can be affected by artifact voltages contributing towards $V_{mix}$,[25,30,31] we use these analyses to primarily gain a qualitative sense of the in-plane and out-of-plane SOT components, and then use the more reliable method of *dc* bias linewidth modulation for SOT evaluation, as discussed below.

To more accurately characterize the SOT efficiency, dc bias dependent STFMR measurements are done to estimate the effective damping-like torque efficiency,[22,25,32] and representative curves for different values of dc current ($I_{dc}$) for TaIrTe4(90 nm)/Py(6 nm) are shown in Fig. **3c**. The resonance linewidht, $\Delta H$, is subsequently extracted for different $I_{dc}$ values, and Fig. **3d** and **3e** show the resulting $\Delta H$ vs. $I_{dc}$ plots for STFMR devices with TaIrTe4(90 nm)/Py(6 nm) and TaIrTe4(64 nm)/Py(5 nm), respectively. The slope [$\delta\Delta H/\delta(I_{dc})$] of linearly fitted $\Delta H$ vs $I_{dc}$ data indicates the strength of SOT and we extract,[22,32]

$$\xi_{DL}^{eff} = \frac{2e}{\hbar} \frac{(H_a+0.5M_{eff})\mu_0 M_S t_{FM}}{\sin\phi} \frac{\gamma}{2\pi f} \frac{\delta\Delta H}{\delta(I_{dc,TaIrTe4})} A_C \qquad (2)$$

with $\phi$ the azimuthal angle between $I_{dc}$ and $\mu_0 H_a$, $M_S = 6.4\times10^5$ A/m the saturation magnetization of the Py layer,[22,33] $t_{FM}$ the thickness of the Py layer, $\frac{\gamma}{2\pi}$ the effective gyromagnetic ratio of Py, $e$ the elementary charge, $\hbar$ the reduced Planck's constant, $I_{dc,TaIrTe4}$ the current in the TaIrTe4 layer, and $A_C$ the cross-sectional area of the STFMR microbars. The $I_{dc,TaIrTe4}$ value is estimated by using the measured resistance of the device and known resistance of the Py layer ($R_{Py}$=329 Ω).[27,28] The effective damping-like torque efficiencies of 0.98±0.21 and 1.18±0.54 are obtained for TaIrTe4 devices of 64, and 90 nm thicknesses, respectively. The effective spin Hall conductivity ($\sigma_{SHC} = \sigma_c \xi_{DL}^{eff}$) values are evaluated to be (1.91±0.42)×10⁴ ($\hbar$/2e) (Ωm)⁻¹ and (7.31±3.30)×10⁴ ($\hbar$/2e) (Ωm)⁻¹ for TaIrTe4 device of 64, and 90 nm, respectively.

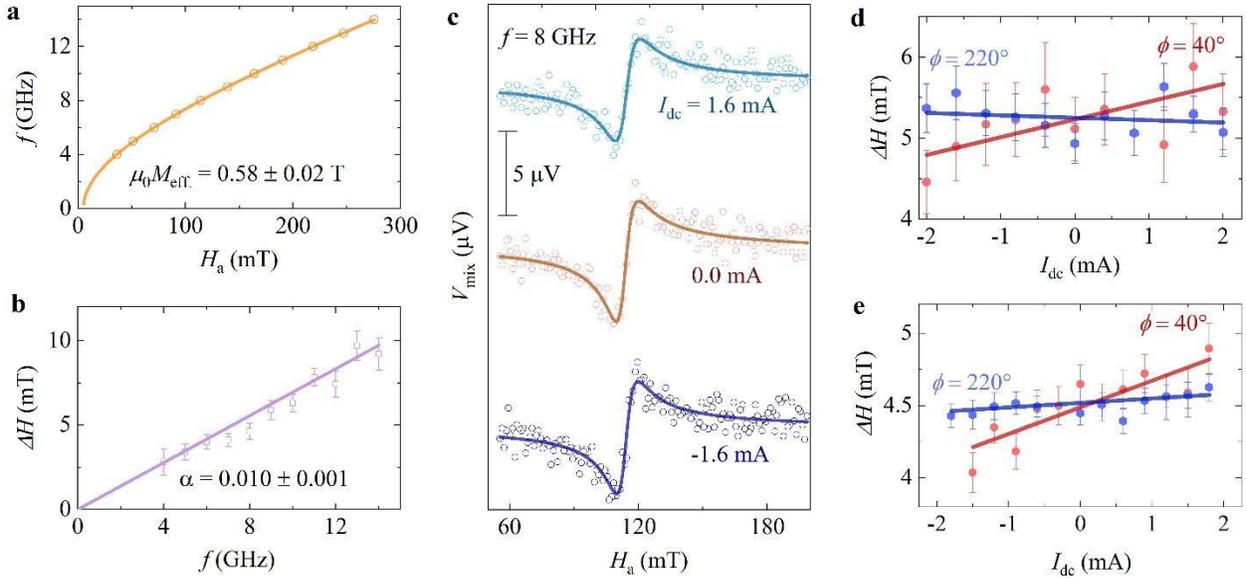

**Figure 3**. Evaluation of effective spin orbit torque efficiency from *dc* bias dependence STFMR measurements. **a,b.** Ferromagnetic resonance frequency *f vs.* resonance field $H_R$ and ferromagnetic resonance linewidth *ΔH vs. f* for TaIrTe$_4$(90 nm)/Py(6 nm) device, respectively. **c,** STFMR curves with different values of *dc* current ($I_{dc}$) at 8 GHz, and **d,** *ΔH* vs. $I_{dc}$ for TaIrTe$_4$ (90 nm)/Py (6 nm) device, respectively. **e,** *ΔH versus* $I_{dc}$ for TaIrTe$_4$(64 nm)/Py (5 nm) device at 7 GHz. In **a-b,** and **d-e,** solid symbols show the extracted values after fitting the experimental data to Eq. 1 and solid lines are fit to the obtained data. In **c**, solid symbols show the experimental data points and solid lines are fit to the data using Eq. 1.

As the amplitude of V$_{mix}$ changes significantly with the sign of applied magnetic field in most of the devices (when $\phi$ is varied from 40° to 220°). This behavior of V$_{mix}$ with $\phi$ indicates the presence of unconventional out-of-plane SOT in this system. To analyze this phenomenon in more detail, we performed angular dependent STFMR measurements, where representative curves for sample TaIrTe$_4$(120 nm)/Py(6 nm) are shown in Fig. **4a**. The *S* and *A* values are extracted for different $\phi$ values and their angular dependence is plotted in Fig. **4b** and **4c**, respectively. In general, the spin current that generates torque has components along all three *x*, *y* and *z* axes, thus the allowed angular dependencies for the coefficients *S* and *A* are,[3,22,34,35]

$$S = S_{DL}^{Y} \cos\phi \sin 2\phi + S_{DL}^{X} \sin\phi \sin 2\phi + S_{FL}^{Z} \sin 2\phi \qquad (3)$$

$$A = A_{FL}^{Y} \cos\phi \sin 2\phi + A_{FL}^{X} \sin\phi \sin 2\phi + A_{DL}^{Z} \sin 2\varphi \qquad (4)$$

where $S_{DL}^{Y}$, $S_{DL}^{X}$, and $A_{DL}^{Z}$ are the coefficients for the damping-like torque with spin polarizations along *x*, *y*, and *z*, respectively; $A_{FL}^{Y}$, $A_{FL}^{X}$, and $S_{FL}^{Z}$ are the corresponding field-like torques. The angular dependence *S* and *A* is fitted using Eqs. (3) and (4) and the obtained parameters are given in Supplementary Table 1. While fits of the symmetric component *S* to Eq. 3 show moderate agreement, the extracted parameters are only required for the *y*- and *x*-polarized spin-current and do not contribute to the *z*-polarized spin-current estimation (see Eqs. 5-7).

By considering that $A_{FL}^{Y}$ is due to the Oersted field alone, the amplitudes of the damping-like torque efficiencies per unit current density in TaIrTe$_4$ layer can be defined as,[22,35]

$$\xi_{DL}^{X} = \frac{S_{DL}^{X}}{A_{FL}^{Y}} \frac{e\mu_0 M_s t_{TaIrTe4} t_{FM}}{\hbar} \sqrt{1 + \left(\frac{M_{eff}}{H_a}\right)} \qquad (5)$$

$$\xi_{DL}^{Y} = \frac{S_{DL}^{Y}}{A_{FL}^{Y}} \frac{e\mu_0 M_s t_{TaIrTe4} t_{FM}}{\hbar} \sqrt{1 + \left(\frac{M_{eff}}{H_a}\right)} \qquad (6)$$

$$\xi_{DL}^{Z} = \frac{A_{DL}^{Z}}{A_{FL}^{Y}} \frac{e\mu_0 M_s t_{TaIrTe4} t_{FM}}{\hbar} \qquad (7)$$

$\xi_{DL}^X$, $\xi_{DL}^Y$, and $\xi_{DL}^Z$ values of 0.04, 0.08, and 0.11 are obtained for the $t_{TaIrTe4} = 120\ nm$ sample. The spin Hall conductivity ($\sigma_{SHC}^k = \sigma_c \xi_{DL}^k$) can be evaluated using the electrical conductivity ($\sigma_c$) obtained from the devices (after considering the electrical resistivity of Py, $\rho_{Py}$= 47 μΩcm), and $\sigma_{SHC}^Z = (4.05\pm0.23)\times10^4$ (ℏ/2e) (Ωm)$^{-1}$, $\sigma_{SHC}^Y = 2.87\times10^4$ (ℏ/2e) (Ωm)$^{-1}$, $\sigma_{SHC}^X=1.43\times10^4$ (ℏ/2e) (Ωm)$^{-1}$ are obtained for the 120 nm TaIrTe4 sample. To investigate whether non-uniformity in the TaIrTe4 thickness makes any difference to the SOT efficiency, STFMR devices were made on such flakes (see Supplementary Fig. 4). Interestingly, $\xi_{DL}^Z$ shows sign reversal and a value of -0.07 is obtained for this device where the average thickness of TaIrTe4 is 64 nm.

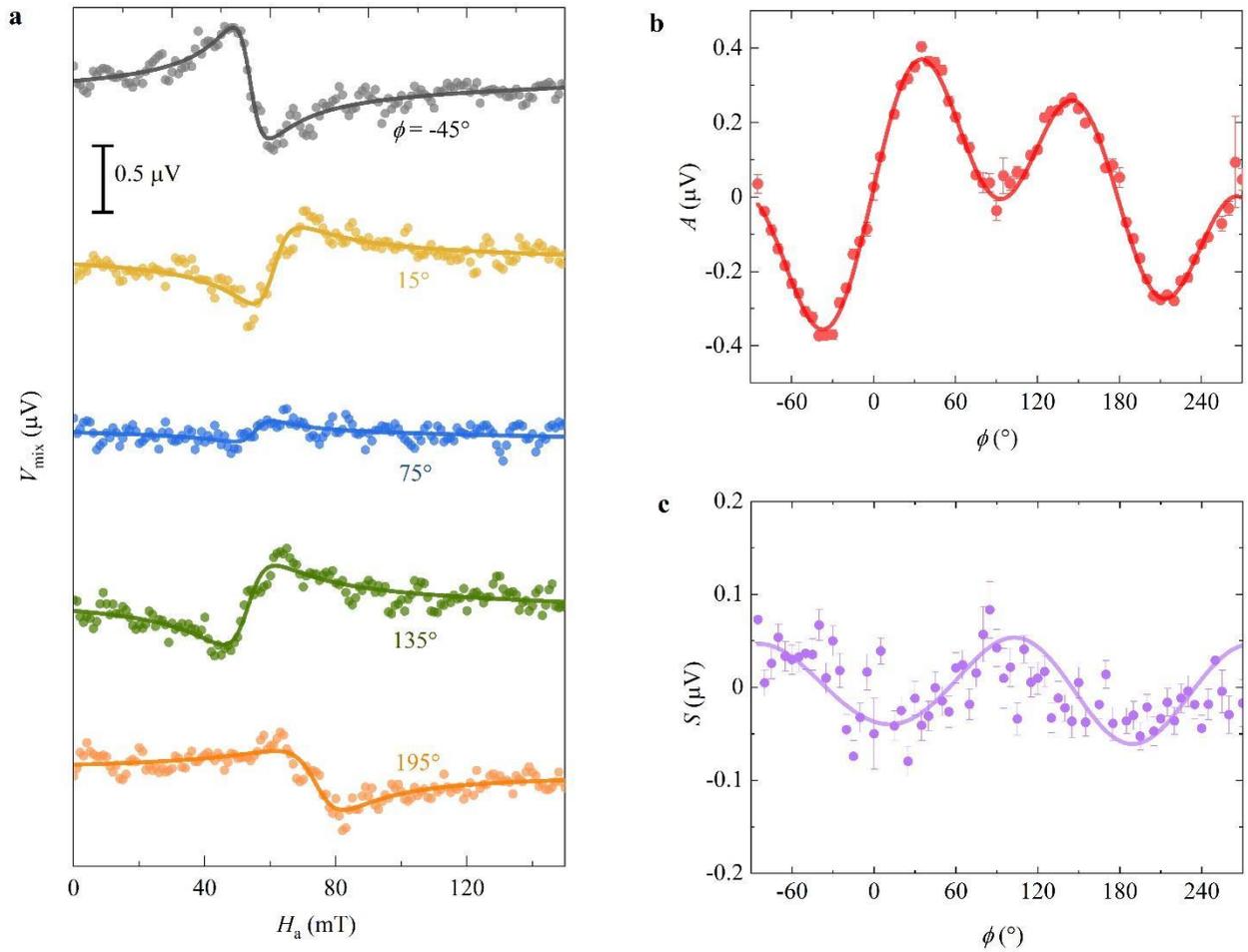

**Figure 4.** Unconventional spin-orbit torque in TaIrTe4 from angular STFMR measurements. **a** The representative STFMR curves at different values of in-plane magnetic field angle $\phi$ for the frequency of 6 GHz. $\phi$ values are shown in the figure. **b,c,** Antisymmetric (*A*) symmetric (*S*) resonance amplitude versus $\phi$. In both **b** and **c**, solid symbols represent the *A* and *S* values extracted after fitting the experimental data to equation (1), and solid lines are fit to the obtained data using Eqs. (3) and (4), respectively. The measurements are performed on the TaIrTe4(120 nm)/Py(6 nm) STFMR device.

**Summary and outlook**

We have demonstrated a large unconventional out-of-plane SOT efficiency in the Weyl semimetal candidate TaIrTe$_4$ at room temperature. Using the STFMR measurements of TaIrTe$_4$/Py heterostructure devices, we observed a large damping-like torque due to efficient charge-to-spin conversion. From angular dependent STFMR measurements, the evaluated out-of-plane damping-like torque and the spin Hall conductivity (SHC) $\sigma_{SHC}^{Z}$ and $\sigma_{SHC}$ values are found to be an *order of magnitude* higher than many transition metal dichalcogenides (TMDs)[11,16,36] including WTe$_2$,[3,5] and comparable to the conventional SOT materials such as Bi$_2$Se$_3$,[37] and heavy metal Pt.[22] In contrast to conventional materials, TaIrTe$_4$ also exhibits a very large out-of-plane SHC $\sigma_{SHC}^{Z}$. Utilization of such topological quantum materials with lower crystal symmetries and novel spin textures are therefore suitable for obtaining large charge-to-spin conversion efficiencies with unconventional out-of-plane SOT components. The antisymmetric SOT component in TaIrTe$_4$ has the potential to realize spintronic technologies by enabling efficient manipulation of magnetic materials with perpendicular magnetic anisotropy, leading to high-density, faster, and energy-efficient memory, logic and communication technologies.

*Note*: After submitting the manuscript we came across a recently published paper on TaIrTe$_4$.[39]

**Methods**

**Single crystal growth:** TaIrTe$_4$ crystals were grown by evaporating tellurium from a tellurium-iridium-tantalum melt. The temperature of the melt and crystal growth was 850°C. The temperature at which tellurium condensed was 720°C.[38] The chemical composition was studied with a digital scanning electronic microscope TESCAN Vega II XMU with energy dispersive microanalysis system INCA Energy 450/XT (20 kV). The EDX analysis showed that the approximate chemical composition of the samples was close to stoichiometric. No impurity elements and phases were found, see Supplementary Fig. 1 for details.

**Device fabrication:** The samples were prepared by mechanically exfoliating nanolayers of TaIrTe$_4$ crystals onto a SiO$_2$/Si wafer by the Scotch tape method inside a glove box. To make TaIrTe$_4$/Ni$_{80}$Fe$_{20}$ heterostructures, the samples were quickly transferred from the glove box to the high vacuum sputtering chamber to deposit Py(5-6 nm)/Al$_2$O$_3$(4 nm) layers. Ni$_{80}$Fe$_{20}$ is known as permalloy (Py). The sample surface was bias sputter cleaned for 20 sec with low energy Ar ion, followed by the deposition of 6 nm Py and 4 nm Al$_2$O$_3$ layers using the dc and rf magnetron sputtering methods, respectively. The TaIrTe$_4$/Py heterostructures were patterned into the rectangular microstrips for spin-torque ferromagnetic resonance measurements (ST-FMR) using the electron-beam lithography and Ar ion milling using the negative e-beam resist as the etching mask. Laser writer lithography was used to prepare the top co-planar waveguide contacts for electrical measurements, followed by the lift-off process of 200 nm of copper (Cu) and 20 nm of platinum (Pt).

**Anisotropic magnetoresistance measurements:** In-plane angular dependence anisotropic magnetoresistance (AMR) measurements were performed on ST-FMR microbars using a rotatable projected vector field magnet with a fixed applied magnetic field of 0.1 T and applied dc current of 0.5 mA. The resistance of the devices was measured while rotating the magnetic field in-plane.

**Spin-orbit torque ferromagnetic resonance (STFMR) measurements:** STFMR measurements were performed at room temperature on the microbar devices to estimate the spin-orbit torque (SOT) efficiency and other magneto dynamical parameters. The measurements for effective SOT analysis were performed with a fixed in-plane angle $\phi=40°$, the radio-frequency (rf) current modulated at 98.76 Hz was applied to the device through a high-frequency bias-T at a fixed frequency (ranging over 3-14 GHz) with an input rf power P=14 dBm. The in-plane angle $\phi$ dependence ST-FMR measurements were performed using a rotatable projected field magnet with $\phi = 0$-$360°$ (step of 5°) at a fixed frequency.

## Acknowledgments

European Union Graphene Flagship (Core 3, No. 881603), 2D TECH VINNOVA center (No. 2019-00068), Swedish Research Council (VR) grant (No. 2021–04821), FLAG-ERA project 2DSOTECH (VR No. 2021-05925), KAW – Wallenberg initiative on Materials Science for a Sustainable World (WISE), Graphene Center, AoA Nano, Chalmers-Max IV collaboration grant, and AoA Materials program at Chalmers University of Technology. MAH acknowledge support from the VR starting grant 2018-05339 and Wallenberg Foundation (Grant No. 2022.0079). We acknowledge the help of staff at the Quantum Device Physics and Nanofabrication laboratory at Chalmers.

## Data availability

The data that support the findings of this study are available from the corresponding authors on a reasonable request.

## Corresponding author

Correspondence to Saroj P. Dash (saroj.dash@chalmers.se), Lakhan Bainsla (lakhan.bainsla@iitrpr.ac.in)

**Contributions**

L.B. fabricated and characterized the devices. B.Z. A.M.H., L.S. and N.B. helped in the device fabrication and measurements. M.A.H. grew the TaIrTe$_4$ crystals, L.B. J.A. and S.P.D. analyzed and interpreted the experimental data and wrote the manuscript with input from all the authors. S.P.D. coordinated and supervised the project.

**Competing interests**

The authors declare no competing interests.

*References*

1. Shao, Q. *et al.* Roadmap of Spin–Orbit Torques. *IEEE Trans. Magn.* **57**, 1–39 (2021).
2. Manchon, A. *et al.* Current-induced spin-orbit torques in ferromagnetic and antiferromagnetic systems. *Rev. Mod. Phys.* **91**, 035004 (2019).
3. MacNeill, D. *et al.* Control of spin–orbit torques through crystal symmetry in WTe2/ferromagnet bilayers. *Nat. Phys.* **13**, 300–305 (2016).
4. MacNeill, D. *et al.* Thickness dependence of spin-orbit torques generated by WTe2. *Phys. Rev. B Condens. Matter* **96**, (2017).
5. Shi, S. *et al.* All-electric magnetization switching and Dzyaloshinskii–Moriya interaction in WTe2/ferromagnet heterostructures. *Nat. Nanotechnol.* **14**, 945–949 (2019).
6. Zhao, B. *et al.* Unconventional charge-spin conversion in Weyl-semimetal WTe2. *Adv. Mater.* **32**, e2000818 (2020).
7. Yang, H. *et al.* Two-dimensional materials prospects for non-volatile spintronic memories. *Nature* **606**, 663–673 (2022).
8. Wieder, B. J. *et al.* Topological materials discovery from crystal symmetry. *Nature Reviews Materials* **7**, 196–216 (2021).
9. Kao, I.-H. *et al.* Deterministic switching of a perpendicularly polarized magnet using unconventional spin–orbit torques in WTe2. *Nat. Mater.* **21**, 1029–1034 (2022).
10. Shin, I. *et al.* Spin-Orbit Torque Switching in an All-Van der Waals Heterostructure. *Adv. Mater.* **34**, e2101730 (2022).
11. Liang, S. *et al.* Spin-orbit torque magnetization switching in MoTe2 /Permalloy heterostructures. *Adv. Mater.* **32**, e2002799 (2020).
12. Vila, M. *et al.* Low-symmetry topological materials for large charge-to-spin interconversion: The case of transition metal dichalcogenide monolayers. *Phys. Rev. Res.* **3**, 043230 (2021).
13. Liu, L. *et al.* Symmetry-dependent field-free switching of perpendicular magnetization. *Nat. Nanotechnol.* **16**, 277–282 (2021).
14. Kurebayashi, H., Garcia, J. H., Khan, S., Sinova, J. & Roche, S. Magnetism, symmetry and spin transport in van der Waals layered systems. *Nature Reviews Physics* **4**, 150–166 (2022).
15. Stiehl, G. M. *et al.* Current-Induced Torques with Dresselhaus Symmetry Due to Resistance Anisotropy in 2D Materials. *ACS Nano* **13**, 2599–2605 (2019).
16. Guimarães, M. H. D., Stiehl, G. M., MacNeill, D., Reynolds, N. D. & Ralph, D. C. Spin-Orbit Torques in NbSe2/Permalloy Bilayers. *Nano Lett.* **18**, 1311–1316 (2018).
17. Ma, J. *et al.* Nonlinear photoresponse of type-II Weyl semimetals. *Nat. Mater.* **18**, 476–481 (2019).

18. Jian, Y. *et al.* Transport signatures of temperature-induced chemical potential shift and Lifshitz transition in layered type-II Weyl semimetal TaIrTe4. *2D Mater.* **8**, 015020 (2020).
19. Xing, Y. *et al.* Surface superconductivity in the type II Weyl semimetal TaIrTe4. *Natl Sci Rev* **7**, 579–587 (2020).
20. Zhuo, X. *et al.* Dynamical evolution of anisotropic response of type-II Weyl semimetal TaIrTe4 under ultrafast photoexcitation. *Light Sci Appl* **10**, 101 (2021).
21. Kumar, D. *et al.* Room-temperature nonlinear Hall effect and wireless radiofrequency rectification in Weyl semimetal TaIrTe4. *Nat. Nanotechnol.* **16**, 421–425 (2021).
22. Liu, L., Moriyama, T., Ralph, D. C. & Buhrman, R. A. Spin-torque ferromagnetic resonance induced by the spin Hall effect. *Phys. Rev. Lett.* **106**, 036601 (2011).
23. Tulapurkar, A. A. *et al.* Spin-torque diode effect in magnetic tunnel junctions. *Nature* **438**, 339–342 (2005).
24. Koepernik, K., Kasinathan, D., Efremov, D. V. & Khim, S. : A ternary type-II Weyl semimetal. *Phys. Rev. B: Condens. Matter Mater. Phys.* (2016) doi:10.1103/PhysRevB.93.201101.
25. Demasius, K.-U. *et al.* Enhanced spin–orbit torques by oxygen incorporation in tungsten films. *Nat. Commun.* **7**, 1–7 (2016).
26. Bainsla, L. *et al.* Ultrathin ferrimagnetic GdFeCo films with low damping. *Adv. Funct. Mater.* **32**, 2111693 (2022).
27. Haidar, M. *et al.* Compositional effect on auto-oscillation behavior of $Ni_{100-x}Fe_x$/Pt spin Hall nano-oscillators. *arXiv [cond-mat.mtrl-sci]* (2020).
28. Haidar, M. *et al.* A single layer spin-orbit torque nano-oscillator. *Nat. Commun.* **10**, 2362 (2019).
29. Shi, S. *et al.* Observation of the out-of-plane polarized spin current from CVD grown $WTe_2$. *Adv. Quantum Technol.* 2100038 (2021).
30. Tserkovnyak, Y., Brataas, A. & Bauer, G. E. W. Enhanced gilbert damping in thin ferromagnetic films. *Phys. Rev. Lett.* **88**, 117601 (2002).
31. Saitoh, E., Ueda, M., Miyajima, H. & Tatara, G. Conversion of spin current into charge current at room temperature: Inverse spin-Hall effect. *Appl. Phys. Lett.* **88**, 182509 (2006).
32. Behera, N. *et al.* Energy-efficient $W_{100-x}Ta_x$/ co-Fe-B/MgO spin hall nano-oscillators. *Phys. Rev. Appl.* **18**, (2022).
33. Krivorotov, I. N. *et al.* Time-domain measurements of nanomagnet dynamics driven by spin-transfer torques. *Science* **307**, 228–231 (2005).
34. Fang, D. *et al.* Spin–orbit-driven ferromagnetic resonance. *Nat. Nanotechnol.* **6**, 413–417 (2011).
35. Bose, A. *et al.* Tilted spin current generated by the collinear antiferromagnet ruthenium dioxide. *Nature Electronics* **5**, 267–274 (2022).
36. Safeer, C. K. *et al.* Room-temperature spin Hall effect in graphene/MoS2 van der Waals heterostructures. *Nano Lett.* **19**, 1074–1082 (2019).
37. Han, J. *et al.* Room-Temperature Spin-Orbit Torque Switching Induced by a Topological Insulator. *Phys. Rev. Lett.* **119**, 077702 (2017).
38. Chareev, D. A. *et al.* Growth of transition-metal dichalcogenides by solvent evaporation technique. *Cryst. Growth Des.* **20**, 6930–6938 (2020).
39. Liu, Y. *et al.* Field-free switching of perpendicular magnetization at room temperature using out-of-plane spins from TaIrTe4. *Nature Electronics* 1–7 (2023). https://doi.org/10.1038/s41928-023-01039-2